\documentclass{article}
\title{
\bf A new Lax pair 
for the sixth Painlev\'e equation
associated with $\widehat{\mathfrak{so}}(8)$
}
\author{Masatoshi NOUMI and Yasuhiko YAMADA\cr
{\normalsize Department of Mathematics, Kobe University}
\medskip
\cr
{\normalsize \it Dedicated to Professor Mitsuo Morimoto on his 
sixtieth birthday}
}
\date{}
\usepackage{amssymb}
\newcommand{\br}[1]{\langle #1\rangle}
\newcommand{\pbr}[1]{\{#1\}}
\newcommand{\BC}{\mathbb{C}}
\newcommand{\BZ}{\mathbb{Z}}
\newcommand{\BP}{\mathbb{P}}
\newcommand{\BS}{\mbox{\boldmath$S$}}
\newcommand{\CK}{\mathcal{K}}

\newcommand{\CM}{\mathcal{M}}
\newcommand{\ba}{\mbox{\boldmath$a$}}
\newcommand{\be}{\mbox{\boldmath$\varepsilon$}}
\newcommand{\bu}{\mbox{\boldmath$u$}}

\newcommand{\Psix}{P_{\mbox{\scriptsize\sc VI}}}
\newcommand{\HSix}{H_{\mbox{\normalsize\sc VI}}}
\newcommand{\Hsix}{H_{\mbox{\scriptsize\sc VI}}}
\newcommand{\dfrac}[2]{{\displaystyle\frac{#1}{#2}}}

\newcommand{\comment}[1]{}
\newcommand{\eqref}[1]{$($\ref{#1}$)$}
\newcommand{\Res}{\mbox{\rm Res}}
\newcommand{\Aut}{\mbox{\rm Aut}}
\newcommand{\diag}[1]{\mbox{\rm diag}(#1)}

\newtheorem{theorem}{Theorem}[section]

\newtheorem{remark}[theorem]{Remark}
\newcommand{\BTqp}{
\begin{table}
\caption{\label{tbl:BTqp}B\"acklund transformations for $\Hsix$} 
\center{
\renewcommand{\arraystretch}{1.6}
\arraycolsep=2.0pt
$\begin{array}{|c||ccccc|cc|}
\hline
& \alpha_0 & \alpha_1 & \alpha_2 & \alpha_3 & \alpha_4 
& q & p\\
\hline\hline
\ s_0\  &-\alpha_0 &\alpha_1& \alpha_2+\alpha_0 & \alpha_3 &\alpha_4
& q & p-\frac{\alpha_0}{q-t} \\
s_1 &\alpha_0 &-\alpha_1& \alpha_2+\alpha_1 & \alpha_3 &\alpha_4
& q & p\\
s_2 &\ \alpha_0+\alpha_2 &\alpha_1+\alpha_2&-\alpha_2 
& \alpha_3+\alpha_2 &\alpha_4+\alpha_2\ 
& \ q+\frac{\alpha_2}{p} & p\\
s_3 &\alpha_0 &\alpha_1& \alpha_2+\alpha_3 & -\alpha_3 &\alpha_4
& q & p-\frac{\alpha_3}{q-1} \\
s_4 &\alpha_0 &\alpha_1& \alpha_2+\alpha_4 & \alpha_3 &-\alpha_4
& q & p-\frac{\alpha_4}{q}\\
\hline
r_1 &\alpha_1 & \alpha_0 & \alpha_2 & \alpha_4 & \alpha_3
&\frac{t(q-1)}{q-t}&-\frac{(q-t)((q-t)p+\alpha_2)}{t(t-1)}
\\
r_3 &\alpha_3 & \alpha_4 & \alpha_2 & \alpha_0 & \alpha_1
&\frac{t}{q}&-\frac{q(qp+\alpha_2)}{t}
\\
r_4 &\alpha_4 & \alpha_3 & \alpha_2 & \alpha_1 & \alpha_3
&\frac{q-t}{q-1}&\frac{(q-1)((q-1)p+\alpha_2)}{t-1}
\\
\hline
\end{array}$
}
\end{table}
}
\begin{document}
\maketitle 
\section*{Introduction}
In this article, we propose a new representation of Lax type  
for the sixth Painlev\'e equation.  
This representation, formulated in the framework 
of the loop algebra $\mathfrak{so}(8)[z,z^{-1}]$ 
of type $D^{(1)}_4$, provides a 
natural explanation of the affine Weyl group symmetry 
of $\Psix$.  
After recalling a standard derivation of $\Psix$, we describe 
in Section 2 fundamental B\"acklund transformations for $\Psix$. 
In Section 3, we present our Lax 
pair for $\Psix$ associated with 
$\mathfrak{so}(8)[z,z^{-1}]$, and 
explain how the B\"acklund transformations 
arise from the linear problem.  
For the general background on Painlev\'e equations, 
we refer the reader to \cite{C}. 

The authors would like to thank Professor Kanehisa Takasaki 
for valuable discussions in the early stage of this work. 

\section{The sixth Painlev\'e equation}
The sixth Painlev\'e equation is the following 
nonlinear ordinary differential equation of second order 
for the unknown function $y=y(t)$: 
\begin{equation}\label{eq:P6}
\arraycolsep=2pt
\begin{array}{ll}\smallskip
y''=&\dfrac{1}{2}
\left(\dfrac{1}{y}+\dfrac{1}{y-1}+\dfrac{1}{y-t}\right)(y')^2
-\left(
\dfrac{1}{t}+
\dfrac{1}{t-1}+
\dfrac{1}{y-t}
\right)y' 
\\
&+
\dfrac{y(y-1)(y-t)}{t^2(t-1)^2}
\left(
\alpha+
\beta\dfrac{t}{y^2}
+\gamma\dfrac{t-1}{(y-1)^2}
+\delta\dfrac{t(t-1)}{(y-t)^2}
\right),
\end{array}\end{equation}
where $'$ stands for the derivation with respect to the 
independent variable $t$, and 
$\alpha$, $\beta$, $\gamma$, $\delta$ are 
complex parameters.  

A standard way to derive the sixth Painlev\'e equation 
is to employ the monodromy preserving deformation 
of a second order Fuchsian differential equation
on $\BP^1$, with four regular singular points and 
one apparent singularity. 
Consider the system of linear differential equations  
\begin{equation}\label{eq:MP}
\big(\partial_x^2+a_1(x,t)\partial_x +a_2(x,t)\big) u=0,
\quad\partial_t u=\big(b_1(x,t)\partial_x+b_2(x,t)\big) u, 
\end{equation}
for the unknown functions $u=u(x,t)$,
where $\partial_x=\partial/\partial x$ and 
$\partial_t=\partial/\partial t$. 
  We assume 
that the coefficients $a_j(x,t)$ and $b_j(x,t)$ are 
rational functions in $x$, depending holomorphically on $t$,
and that the first equation is Fuchsian with Riemann scheme
\begin{equation}
\left\{
\begin{array}{ccccc}\smallskip
x=0 & x=1 & x=t & x=q & x=\infty \\
0 & 0 & 0 & 0 & \rho \\
\kappa_0 & \kappa_1 & \kappa_t & 2 & \kappa_\infty+\rho
\end{array}
\right\}
\end{equation}
with respect to the variable $x$. 
In this scheme,  
$\kappa_0$, $\kappa_1$, $\kappa_t$, $\kappa_\infty$ and $\rho$ 
are generic complex parameters subject to the {\em Fuchs relation}
\begin{equation}
\kappa_0+\kappa_1+\kappa_t+\kappa_\infty+2 \rho=1. 
\end{equation}
We also assume that the singularity $x=q$, which may depend on $t$, is 
{\em non-logarithmic}. 
Under these assumptions, the coefficients $a_1(x,t)$, 
$a_2(x,t)$ are expressed in the form 
\begin{equation}
\arraycolsep=2pt
\begin{array}{llllllll}\smallskip
a_1(x,t)&=\dfrac{1-\kappa_0}{x}+
\dfrac{1-\kappa_1}{x-1}+
\dfrac{1-\kappa_t}{x-t}-
\dfrac{1}{x-q},
\\
a_2(x,t)&=\dfrac{1}
{x(x-1)}\left\{
-\dfrac{t(t-1) H}{x-t} +\dfrac{q(q-1) p}{x-q}+\rho(\kappa_\infty+\rho)
\right\},
\end{array}\end{equation}
respectively, where 
\begin{equation}
p=\Res_{x=q}(a_2(x,t)dx),\quad H=-\Res_{x=t}(a_2(x,t)dx). 
\end{equation}
Furthermore, the coefficient $H$ is determined 
as a polynomial in $(q,p)$ with coefficients 
in $\BC(t)$; explicitly, it is given by 
\begin{equation}
\label{eq:Ham}
\arraycolsep=2pt
\begin{array}{llllllll}
H=\dfrac{1}{t(t-1)}\big[&
p^2 q(q-1)(q-t) -p \big\{\kappa_0 (q-1)(q-t)
\\
&+\kappa_1 q(q-t)
+(\kappa_t-1)q(q-1)\big\}+\rho(\kappa_\infty+\rho) (q-t)\big]. 
\end{array}\end{equation}
The compatibility condition of the linear differential 
system \eqref{eq:MP} then 
turns out to be expressed as the Hamiltonian system 
\begin{equation}\label{eq:H6}
\Hsix:\qquad
\partial_t(q)=\frac{\partial H}{\partial p},
\quad
\partial_t(p)=-\frac{\partial H}{\partial q}, 
\end{equation} 
with polynomial Hamiltonian $H$ in \eqref{eq:Ham}; 
namely,
\begin{equation}
\arraycolsep=2pt
\begin{array}{lll}
\smallskip
t(t-1)\partial_t(q)=&
2 p q(q-1)(q-t) -\big\{\kappa_0 (q-1)(q-t) \\ 
\smallskip
&+\,\kappa_1 q(q-t)
+(\kappa_t-1)q(q-1)\big\}\\
\smallskip
t(t-1)\partial_t(p)=&-p^2(3q^2-2(1+t)q+t)
+p\big\{2(\kappa_0+\kappa_1+\kappa_t-1)q
\\
&-\,\kappa_0(1+t)-\kappa_1 t-\kappa_t+1\big\}
-\rho(\kappa_\infty+\rho). 
\end{array}
\end{equation}
This system of nonlinear equations is in fact equivalent to 
the sixth Painlev\'e equation $\Psix$ \eqref{eq:P6}
for $y=q$, with parameters
\begin{equation}
\alpha=\frac{\kappa_\infty^2}{2},
\quad
\beta=-\frac{\kappa_0^2}{2},
\quad
\gamma=\frac{\kappa_1^2}{2},
\quad
\delta=\frac{1-\kappa_t^2}{2}. 
\end{equation}
We remark that, 
in place of  \eqref{eq:MP}, 
one can naturally make use of the Schlesinger system of 
rank two, with regular singular points $x=0,1,t,\infty$. 

\section{Discrete symmetry of $\HSix$}
It is known that the sixth Painlev\'e equation admits 
a group of B\"acklund transformations which is 
isomorphic to the (extended) affine Weyl group of 
type $D^{(1)}_4$ (see \cite{O}, for instance).

In describing the B\"acklund transformations for $\Hsix$,  
it is convenient to use the  parameters 
\begin{equation}
\alpha_0=\kappa_t,\quad 
\alpha_1=\kappa_\infty,\quad
\alpha_2=\rho,\quad
\alpha_3=\kappa_1,\quad
\alpha_4=\kappa_0
\end{equation}
with linear relation 
$\alpha_0+\alpha_1+2\alpha_2+\alpha_3+\alpha_4=1$, 
so that
\begin{equation}
\arraycolsep=2pt
\begin{array}{llllllll}\smallskip
t(t-1)H=&
p^2 q(q-1)(q-t) -p \big\{
(\alpha_0-1)q(q-1)
\\
&
+\alpha_3 q(q-t)
+\alpha_4 (q-1)(q-t)
\big\}+\alpha_2(\alpha_1+\alpha_2) (q-t). 
\end{array}\end{equation}
In the following, we identify the parameter space for 
$\Hsix$ with the affine space 
$V=\BC^4$ with canonical coordinates 
$\varepsilon=(\varepsilon_1,\varepsilon_2,
\varepsilon_3,\varepsilon_4)$, and regard 
$\alpha_j$ as linear functions of $V$ such that 
\begin{equation}
\begin{array}{lll}
\alpha_0=1-\varepsilon_1-\varepsilon_2,\quad
&\alpha_1=\varepsilon_1-\varepsilon_2,\quad
&\alpha_2=\varepsilon_2-\varepsilon_3,\cr
\alpha_3=\varepsilon_3-\varepsilon_4,
&\alpha_4=\varepsilon_3+\varepsilon_4. 
\end{array}
\end{equation}
We identify  $V$ with the Cartan subalgebra 
of the simple Lie algebra $\mathfrak{so}(8)$
(of type $D_4$); 
$\pbr{\varepsilon_1,\ldots,\varepsilon_4}$ is then 
a canonical orthonormal basis of $V^\ast$, and 
$\alpha_0,\alpha_1,\alpha_2,\alpha_3,\alpha_4$ 
are the {\em simple affine roots}. 
Note that the null root $\delta=\alpha_0+\alpha_1+2\alpha_2+\alpha_3+\alpha_4$
is normalized to be the constant function $1$.

By a {\em B\"acklund transformation}, 
we mean a transformation of dependent variables 
and parameters that leaves the system invariant. 
Let us show an example of B\"acklund transformation for  
$\Hsix$.   
Define new variables $\widetilde{q}, \widetilde{p}$
by
\begin{equation}
\widetilde{q}=q,\qquad \widetilde{p}=p-\frac{\alpha_0}{q-t}. 
\end{equation}
Then one can verify directly that,
if the pair $(q,p)$ satisfies the Hamiltonian system \eqref{eq:H6},  
then the pair $(\widetilde{q},\widetilde{p})$ again satisfies 
the same system with parameters 
$\alpha_0$, $\alpha_2$ replaced by $-\alpha_0$, $\alpha_2+\alpha_0$,
respectively; 
we refer to this B\"acklund transformation as $s_0$. 
To be more precise, let us consider the field of rational functions 
\begin{equation} 
\CK=\BC(\alpha_1,\alpha_2,\alpha_3,\alpha_4,q,p,t)
\qquad(\alpha_0=1-\alpha_1-2\alpha_2-\alpha_3-\alpha_4),
\end{equation}
and the Hamiltonian vector field 
\begin{equation}
\delta=\frac{\partial H}{\partial p}\frac{\partial\ }{\partial q}
-\frac{\partial H}{\partial q}\frac{\partial\ }{\partial p}
+\frac{\partial\ }{\partial t}
\end{equation} 
acting on $\CK$ as a derivation. 
We regard this {\em differential field\/} $(\CK,\delta)$ as 
representing the Hamiltonian system $\Hsix$. 
We define the automorphism $s_0 : \CK \to \CK$ 
by setting
\begin{equation}
s_0(\alpha_0)=-\alpha_0,\quad
s_0(\alpha_2)=\alpha_2+\alpha_0,
\quad
s_0(\alpha_j)=\alpha_j\quad(j\ne 0,2), 
\end{equation}
and 
\begin{equation}
s_0(q)=q,\quad
s_0(p)=p-\frac{\alpha_0}{q-t},\quad s_0(t)=t. 
\end{equation}
Then one can show that the automorphism $s_0:\CK \to \CK$ 
commutes with the Hamiltonian vector field $\delta$. 
In this sense, a B\"acklund transformation can be defined 
alternatively to be 
an automorphism of the differential field that commutes with 
the derivation.  

Table \ref{tbl:BTqp} is the list of 
{\em fundamental} B\"acklund transformations for $\Hsix$. \BTqp 
We consider two subgroups 
\begin{equation}
W=\br{s_0,s_1,s_2,s_3,s_4}\subset 
\widetilde{W}=\br{s_0,s_1,s_2,s_3,s_4,r_1,r_3,r_4}
\subset \Aut_\delta(\CK)
\end{equation}
of differential automorphisms of $\CK$, 
generated by the B\"acklund transformations in Table \ref{tbl:BTqp}.  
Then it turns out that 
$W$ and $\widetilde{W}$ are 
isomorphic to the affine Weyl group
and the extended affine Weyl group  
of type $D^{(1)}_4$, respectively. 
The B\"acklund transformations $s_i$ ($i=0,1,2,3,4$) 
and $r_i$ ($i=1,3,4$) 
in fact satisfy the fundamental relations 
\begin{equation}
\renewcommand{\arraystretch}{1.4}
\begin{array}{llllllll}
& s_i^2=1&&(i=0,1,2,3,4),
\\
& s_i s_j=s_js_i&&(i,j=0,1,3,4),
\\
& s_i s_2 s_i=s_2 s_i s_2 &&(i=0,1,3,4),
\\
& r_i^2=1 && (i=1,3,4)
\\
& r_ir_j=r_k && (\pbr{i,j,k}=\pbr{1,3,4})
\\
& r_i s_j= s_{\sigma_i(j)} r_i &&
(i=1,3,4;\ j=0,1,2,3,4),
\end{array}\end{equation}
where $\sigma_i$ ($i=1,3,4$) are the permutations defined by 
\begin{equation}
\sigma_1=(01)(34),\quad 
\sigma_3=(03)(14),\quad 
\sigma_4=(04)(13). 
\end{equation} 
We also remark that each element $w\in \widetilde{W}$ defines 
a {\em canonical transformation}: 
\begin{equation}
w(\pbr{\varphi,\psi})=\pbr{w(\varphi),w(\psi)}
\qquad(\varphi,\psi\in \CK), 
\end{equation}
where $\pbr{\,,\,}$ stands for the standard Poisson bracket defined 
by
\begin{equation}
\pbr{\varphi,\psi}=
\frac{\partial \varphi}{\partial p}\frac{\partial \psi}{\partial q}
-\frac{\partial \varphi}{\partial q}\frac{\partial \psi}{\partial p}.
\end{equation}

\begin{remark}\rm 
The fundamental relations for the generators $s_0, s_1, s_2, s_3, s_4$
of the affine Weyl group of type $D^{(1)}_4$ 
is described as follows in terms of the Cartan matrix 
$A=\pmatrix{a_{ij}}_{i,j=0}^4$:
\begin{equation}
\begin{array}{cl}\smallskip
s_i^2=1\quad& (i=0,1,2,3,4)\cr\smallskip
s_is_j=s_js_i& \mbox{if}\quad (a_{ij},a_{ji})=(0,0),\cr
\quad s_is_js_i=s_js_is_j\quad& \mbox{if}\quad (a_{ij},a_{ji})=(-1,-1),
\end{array}
\end{equation}
where 
\begin{equation}
A=
\left[\matrix{
2 & 0 & -1 & 0 & 0 \cr
0 & 2 & -1 & 0 & 0 \cr
-1 & -1 & 2 & -1 & -1 \cr
0 & 0 & -1 & 2 & 0 \cr
0 & 0 & -1 & 0 & 2
}\right].\qquad
\begin{picture}(70,0)(-10,0)
\put(0,20){$\circ$} 
\put(20,0){$\circ$} 
\put(0,-20){$\circ$} 
\put(40,20){$\circ$}
\put(40,-20){$\circ$}
\put(4,21){\line(1,-1){17}}
\put(4,-16){\line(1,1){17}}
\put(41,21){\line(-1,-1){17}}
\put(41,-16){\line(-1,1){17}}
\put(-8,20){\small$0$}
\put(-8,-20){\small$1$}
\put(20,-10){\small$2$}
\put(47,20){\small$3$}
\put(47,-20){\small$4$}
\end{picture}
\end{equation}
The action of $s_i$ on the simple affine roots $\alpha_j$ 
is given by
\begin{equation}
s_i(\alpha_j)=\alpha_j-\alpha_i a_{ij}\qquad
(i,j=0,1,2,3,4).
\end{equation}
Note also that 
$\widetilde{W}$ is isomorphic to the semidirect 
product $W\rtimes \Omega$ of 
$W$ and  
$\Omega=\pbr{1,r_1,r_3,r_4}$
acting on $W$ through 
the permutations $\pbr{1,\sigma_1,\sigma_3,\sigma_4}$ 
of indices 
for the generators $s_j$; $\Omega$ 
is identified with a group of 
diagram automorphisms of the Dynkin diagram of type 
$D^{(1)}_4$.  
If we set 
\begin{equation}
\varphi_0=q-t,\quad
\varphi_1=1,\quad \varphi_2=-p,\quad 
\varphi_3=q-1,\quad \varphi_4=q,
\end{equation}
the B\"acklund transformations $s_i$ are
expressed as 
\begin{equation}
s_i(\varphi_j)=\varphi_{j}+\frac{\alpha_i}{\varphi_i} u_{ij},
\quad u_{ij}=\pbr{\varphi_i,\varphi_j}
\quad(i,j=0,1,2,3,4), 
\end{equation}
consistently with the birational Weyl group actions 
discussed previously in \cite{NY}, \cite{NY2}. 
In this particular case, 
the matrix $U=\pmatrix{u_{ij}}_{i,j=0}^4$ is given by 
\begin{equation}
U=\left[\matrix{
0 & 0 & 1 & 0 & 0 \cr
0 & 0 & 0 & 0 & 0 \cr
-1 & 0 & 0 & -1 & -1 \cr
0 & 0 & 1 & 0 & 0 \cr
0 & 0 & 1 & 0 & 0
}\right]. 
\end{equation} 
Observe that $u_{12}=u_{21}=0$;  this degeneracy is caused 
by the normalization that one of the regular singular points 
of \eqref{eq:MP} is 
placed at $x=\infty$. 
\end{remark}
\begin{remark}\rm 
The extended affine Weyl group $\widetilde{W}$ 
of type $D^{(1)}_4$ is expressed 
as the semidirect product of the weight lattice 
$P$ of type $D_4$ and the Weyl group 
$W(D_4)=\br{s_1,s_2,s_3,s_4}$ acting on $P$:
\begin{equation}
\widetilde{W}\stackrel{\sim}{\leftarrow} P\rtimes W(D_4),\quad
P=\bigoplus_{i=1}^4 \BZ\,\varpi_i,
\end{equation}
where $\varpi_i$ are the fundamental weights of 
type $D_4$  defined by
\begin{equation}
\begin{array}{lllll}\smallskip
\varpi_1=\varepsilon_1,\quad&
\varpi_2=\varepsilon_1+\varepsilon_2,\cr
\varpi_3=\frac{1}{2}(\varepsilon_1+
\varepsilon_2+\varepsilon_3-\varepsilon_4),\quad
&\varpi_4=\frac{1}{2}(\varepsilon_1+
\varepsilon_2+\varepsilon_3+\varepsilon_4). 
\end{array}
\end{equation}
Note also that the weight lattice $P$ is the set of all elements 
\begin{equation}
\lambda=\frac{1}{2}(n_1\varepsilon_1+
n_2\varepsilon_2+
n_3\varepsilon_3+
n_4\varepsilon_4)\qquad (n_1,n_2,n_3,n_4\in\BZ) 
\end{equation}
such that, either all the $n_j$'s are even, or all the $n_j$'s are odd. 
The translations $T_{\varpi_i}$ ($i=1,\dots,4$) corresponding 
to $\varpi_i$ are expressed as
\begin{equation}
\begin{array}{llll}\smallskip
T_{\varpi_1}=r_1 s_1 s_2 s_3 s_4 s_2 s_1, \quad&
T_{\varpi_2}=s_0 s_2 s_1 s_3 s_4 s_2 s_1 s_3 s_4 s_2, \cr
T_{\varpi_3}=r_3 s_3 s_2 s_1 s_4 s_2 s_3,&
T_{\varpi_4}=r_4 s_4 s_2 s_1 s_3 s_2 s_4,
\end{array}
\end{equation}
in terms of the generators $s_j$ and $r_j$. 
These elements transform the simple affine roots 
$\alpha_j$ as follows.
\begin{equation}
\def\arraystretch{1.2}
\begin{array}{|c||ccccc|}
\hline
& \alpha_0 & \alpha_1 & \alpha_2 & \alpha_3 & \alpha_4 \cr
\hline\hline
T_{\varpi_1} & \alpha_0+1 & \alpha_1-1 & \alpha_2 & \alpha_3 & \alpha_4 \cr
T_{\varpi_2} & \alpha_0+2 & \alpha_1 & \alpha_2-1 & \alpha_3 & \alpha_4 \cr
T_{\varpi_3} & \alpha_0+1 & \alpha_1 & \alpha_2 & \alpha_3-1 & \alpha_4 \cr
T_{\varpi_4} & \alpha_0+1 & \alpha_1 & \alpha_2 & \alpha_3 & \alpha_4-1\cr
\hline
\end{array}
\end{equation}
Regarded as automorphisms of 
$\CK=\BC(\alpha_1,\alpha_2,\alpha_3,\alpha_4,q,p,t)$, 
$T_{\varpi_i}$ $(i=1,2,3,4)$ 
provide a commuting family of 
B\"acklund transformations for $\Hsix$, which are 
called {\em Schlesinger transformations} 
(such B\"acklund transformations that act as 
shift operators on the parameter space). 
\end{remark}
\par\medskip
A certain part of this discrete symmetry of $\Hsix$ can be
explained by the monodromy preserving deformation of 
a second order Fuchsian equation \eqref{eq:MP}. 
In fact, each $s_i$ ($i=0,1,3,4$), 
except $s_2$, arises from a simple transformation 
of the unknown function $u=u(x,t)$, and  
each $r_i$ ($i=1,3,4$) from a fractional linear 
transformation of the coordinate $x$. 
This framework does not seem, however, to explain  
the particular B\"acklund transformation 
$s_2$ in Table \ref{tbl:BTqp}, which is essential in 
understanding the whole picture of discrete symmetry of 
$\Hsix$. 
In the next section, we propose a new Lax pair  
for $\Hsix$, from which all the B\"acklund 
transformations in Table \ref{tbl:BTqp} can be 
understood naturally.  

\section{Lax pair associated with $\widehat{\mathfrak{so}}(8)$}

Consider the following system of linear differential equations 
for the column vector 
$\vec{\psi}=(\psi_1,\psi_2,\ldots,\psi_8)^{\mbox{\scriptsize\rm t}}$
of eight unknown functions $\psi_i=\psi_i(z,t)$ ($i=1,2,\ldots,8$): 
\begin{equation}\label{eq:Lax}
(z\partial_z+M)\vec{\psi}=0,\quad
\partial_t \vec{\psi}=B \vec{\psi}, 
\end{equation}
with the compatibility condition
\begin{equation}\label{eq:MBCC}
\big[z\partial_z+M,\partial_t-B\big]=0.
\end{equation}
We assume that the matrices $M$ and $B$ are in the form 
\begin{equation}\label{eq:MB} 
M=\left[\matrix{
\varepsilon_1&1&0&0&0&0&0&0 \cr
0&\varepsilon_2&-p&-1&-1&0&0&0 \cr
0&0&\varepsilon_3&q-1&q&0&0&0\cr
0&0&0&\varepsilon_4&0&-q&1&0\cr
0&0&0&0&-\varepsilon_4&1-q&1&0\cr
-z&0&0&0&0&-\varepsilon_3&p&0\cr
(t-q)z&0&0&0&0&0&-\varepsilon_2&-1\cr
0&(q-t)z&z&0&0&0&0&-\varepsilon_1
}\right]
\end{equation}
and
\begin{equation}
B=\left[\matrix{
u_1&x_1&y_1&0&0&0&0&0\cr
0&u_2&x_2&-y_3&-y_4&0&0&0\cr
0&0&u_3&x_3&x_4&0&0&0\cr
0&0&0&u_4&0&-x_4&y_4&0\cr
0&0&0&0&-u_4&-x_3&y_3&0\cr
0&0&0&0&0&-u_3&-x_2&-y_1\cr
-z&0&0&0&0&0&-u_2&-x_1\cr
0&z&0&0&0&0&0&-u_1
}\right], 
\end{equation}
respectively,
where $\varepsilon_j$ are complex constants,
and the variables $q, p, x_j, y_j$ and $u_j$ are  
functions in $t$. 
As before, we set 
\begin{equation}\label{eq:affrts}
\begin{array}{lllll}\smallskip
\alpha_0=1-\varepsilon_1-\varepsilon_2,\quad
&\alpha_1=\varepsilon_1-\varepsilon_2,\quad
&\alpha_2=\varepsilon_2-\varepsilon_3,\\
\alpha_3=\varepsilon_3-\varepsilon_4,
&\alpha_4=\varepsilon_3+\varepsilon_4.
\end{array}\end{equation}
\begin{theorem}
Under the compatibility condition \eqref{eq:MBCC}, 
the variables $x_j$, $y_j$ and $u_j$ are determined 
uniquely as elements of 
$\CK=\BC(\alpha_1,\alpha_2,\alpha_3,\alpha_4,q,p,t)$.
The compatibility condition is then equivalent to the 
Hamiltonian system $\Hsix$ of the sixth
Painlev\'e equation. 
\end{theorem}
The variables $x_j$, $y_j$ and $u_j$ 
appearing in $B$ are determined explicitly as follows:
\begin{eqnarray}
\renewcommand{\arraystretch}{1.8}
&\begin{array}{lllll}
x_1=\dfrac{q-t}{t(t-1)},\quad 
&x_3=-\dfrac{q}{t}, \quad
&x_4=-\dfrac{q-1}{t-1},
\cr
y_1=\dfrac{1}{t(t-1)},
& y_3=-\dfrac{1}{t},
 & y_4=-\dfrac{1}{t-1},
\end{array}\nonumber\\
&\quad x_2=-\dfrac{(q-t)p+\alpha_1+\alpha_2}{t(t-1)},
\end{eqnarray}
and 
\begin{equation}
\renewcommand{\arraystretch}{1.4}
\begin{array}{lllll}
t(t-1)\,u_1=
-q(q-1)p-\alpha_2 q 
+\frac{\alpha_0-\alpha_1-1}{2}t
-\frac{\alpha_0+\alpha_4-1}{2},\cr
t(t-1)\,u_2=
-q(q-1)p-(\alpha_1+\alpha_2)q 
+\frac{\alpha_0+\alpha_1-1}{2}t
-\frac{\alpha_0+\alpha_4-1}{2},\cr
t(t-1)\,u_3=
(2q-1)(q-t)p+(\alpha_1+2\alpha_2)q 
+\frac{\alpha_3+\alpha_4}{2}t
+\frac{\alpha_0+\alpha_4-1}{2},\cr
t(t-1)\,u_4=
-(q-t)p
+\frac{\alpha_3-\alpha_4}{2}t
+\frac{\alpha_0+\alpha_4-1}{2}.
\end{array}\nonumber
\end{equation} 

We now explain how this representation is 
related to 
the Lie algebra $\mathfrak{so}(8)$ and its 
loop algebra. 
With the notation of matrix units 
$E_{ij}=\pmatrix{\delta_{ia}\delta_{jb}}_{a,b=1}^8$, 
we set 
\begin{equation}
J=\sum_{i=1}^8 E_{i,9-i}. 
\end{equation}
We consider the following realization 
of the simple Lie algebra $\mathfrak{so}(8)$:
\begin{equation}
\mathfrak{so}(8)=\pbr{X\in \mbox{Mat}(8;\BC)\mid 
J X + X^{\mbox{\scriptsize t}}J=0},
\end{equation}
where $X^{\mbox{\scriptsize t}}$ denotes the transposition 
of $X$. 
Let us define the Chevalley generators 
$E_j, H_j, F_j$ ($j=0,1,2,3,4$) 
for the loop 
algebra $\mathfrak{so}(8)[z,z^{-1}]$ by
\begin{equation}
\renewcommand{\arraystretch}{1.4}
\begin{array}{lll}
E_0=z(E_{82}-E_{71}),\quad 
&E_1=E_{12}-E_{78},\quad
&E_2=E_{23}-E_{67},\cr
E_3=E_{34}-E_{56},\quad
&E_4=E_{35}-E_{46}. 
\cr
F_0=z^{-1}(E_{28}-E_{17}),\quad&
F_1=E_{21}-E_{87},\quad& 
F_2=E_{32}-E_{76},\cr
F_3=E_{43}-E_{65},&
F_4=E_{53}-E_{64},
\end{array} 
\end{equation}
and $H_j=[E_j,F_j]$  ($j=0,1,2,3,4$).
For a vector $\ba=(a_1,a_2,a_3,a_4)$ given, we also use the 
notation 
\begin{equation}
H(\ba)=\sum_{i=1}^4  a_i(E_{ii}-E_{9-i,9-i})
\end{equation}
for the corresponding 
element in the Cartan subalgebra of $\mathfrak{so}(8)$, 
so that
\begin{equation}
\begin{array}{lll}\smallskip
H_0=H(-1,-1,0,0),\ \ & 
H_1=H(1,-1,0,0),\ \ &
H_2=H(0,1,-1,0)\cr
H_3=H(0,0,1,-1),&
H_4=H(0,0,1,1). 
\end{array}
\end{equation}
Notice that the two matrices 
$M, B$ belong to a Borel subalgebra 
of the loop algebra $\mathfrak{so}(8)[z,z^{-1}]$;
in fact, they are expressed in the form 
\begin{equation}
\arraycolsep=2pt
\renewcommand{\arraystretch}{1.4}
\begin{array}{ll}
M=&H(\be)+(q-t)E_0+E_1-p E_2+(q-1)E_3+q E_4
\cr
&+[E_0,E_2]+[E_3,E_2]+[E_4,E_2],\cr
B=&H(\bu)
+E_0+x_1 E_1+x_2 E_2+x_3 E_3+x_4 E_4
\cr
&+y_1[E_1,E_2]+y_3[E_3,E_2]+y_4[E_4,E_2],
\end{array}
\end{equation}
where 
$\be=(\varepsilon_1,\varepsilon_2,\varepsilon_3,\varepsilon_4)$
and $\bu=(u_1,u_2,u_3,u_4)$. 
\begin{remark} \rm
The affine Lie algebra 
$\widehat{\mathfrak{so}}(8)$ is realized as a central extension 
of the loop algebra $\mathfrak{so}(8)[z,z^{-1}]$, together with the 
derivation $d=z\partial_z$ (see \cite{K}, for the detail)\,$:$ 
\begin{equation}
\widehat{\mathfrak{so}}(8)
=\mathfrak{so}(8)\otimes\BC[z,z^{-1}]
\oplus \BC c \oplus \BC d,
\end{equation}
where $c$ denotes the canonical central element. 
In this context, the simple affine roots $\alpha_j$ ($j=0,1,2,3,4$) 
are defined as linear functionals of the Cartan subalgebra
\begin{equation}
\mathfrak{h}=\bigoplus_{i=0}^4 \BC h_i\oplus \BC d,
\quad
h_0=H_0\otimes 1+c,\quad h_i=H_i\otimes 1 \quad(i=1,2,3,4)
\end{equation} 
such that 
\begin{equation}
\br{h_i,\alpha_j}=a_{i,j}\quad(i=0,1,2,3,4),
\quad
\br{d,\alpha_j}=\delta_{0,j}
\end{equation}
for $j=0,1,2,3,4$; also they are extended to linear functionals 
on the whole affine Lie algebra $\widehat{\mathfrak{so}}(8)$ 
through the triangular decomposition. 
Note that our Lax pair mentioned above 
is formulated in fact in the framework of 
$\widehat{\mathfrak{so}}(8)/\BC c
=\mathfrak{so}(8)[z,z^{-1}]\oplus \BC d$: 
\begin{equation}
\CM =d+M=z\partial_z+M \in\widehat{\mathfrak{so}}(8)/\BC c. 
\end{equation}
Since $\br{c,\alpha_j}=0$, we can regard $\alpha_j$ 
($j=0,1,2,3,4$) and 
the null root $\delta=\alpha_0+\alpha_1+2\alpha_2+\alpha_3+
\alpha_4$ 
as linear functionals on $\widehat{\mathfrak{so}}(8)/\BC c$.  
Then we have 
\begin{equation}\begin{array}{lllll}\smallskip
\alpha_0(\CM)=1-\varepsilon_1-\varepsilon_2,\ \ 
&\alpha_1(\CM)=\varepsilon_1-\varepsilon_2,\ \ 
&\alpha_2(\CM)=\varepsilon_2-\varepsilon_3,\\
\alpha_3(\CM)=\varepsilon_3-\varepsilon_4,
&\alpha_4(\CM)=\varepsilon_3+\varepsilon_4,
& \ \ \delta(\CM)=1. 
\end{array}\end{equation}
In this sense, our notation for the 
parameters \eqref{eq:affrts} is consistent with that of 
simple roots for $\widehat{\mathfrak{so}}(8)$. 
\end{remark}
\par\medskip
In our framework,  
the B\"acklund transformations for $\Hsix$ are 
obtained as the gauge transformations
\begin{equation}\label{eq:BTLax}
s_k\vec\psi=G_k \vec\psi\quad(k=0,1,2,3,4),
\qquad r_k\vec\psi=\Gamma_k\,\vec\psi\quad(k=1,3,4)
\end{equation}
of the linear problem \eqref{eq:Lax}, 
defined by certain matrices 
$G_k$, $\Gamma_k$ in the loop group of 
\begin{equation}
SO(8)=\pbr{X\in SL(8;\BC) \mid X^{\mbox{\scriptsize t}} J X=J}. 
\end{equation}
The matrices $G_k$ and $\Gamma_k$ will be specified below. 
\begin{theorem}\label{thm:B}
The B\"acklund transformations $s_k$ and $r_k$ for $\Hsix$ 
are recovered from the compatibility conditions 
\begin{equation}
s_k(M)=G_k MG_k^{-1}-z\partial_z(G_k)G_k^{-1},\quad
s_k(B)=G_k BG_k^{-1}+\partial_t(G_k)G_k^{-1},
\end{equation}
and 
\begin{equation}\label{eq:CCR}
r_k(M)=\Gamma_k M\Gamma_k^{-1}-z\partial_z(\Gamma_k)\Gamma_k^{-1},
\quad
r_k(B)=\Gamma_k B\Gamma_k^{-1}+\partial_t(\Gamma_k)\Gamma_k^{-1}. 
\end{equation}
\end{theorem} 
In \eqref{eq:BTLax}, the matrices $G_k$ are determined as  
\begin{equation}
\begin{array}{lll}\smallskip
G_0=1+\dfrac{\alpha_0}{q-t}\,F_0,\quad &
G_1=1+\alpha_1 F_1,\quad &
G_2=1-\dfrac{\alpha_2}{p} F_2,\cr
G_3=1+\dfrac{\alpha_3}{q-1}\,F_3,\quad &
G_4=1+\dfrac{\alpha_4}{q}\,F_4.
\end{array}
\end{equation}
The matrices $\Gamma_1, \Gamma_3$ and $\Gamma_4$ are given explicitly by
\newcommand{\rtt}{\sqrt{t(t-1)}}
\begin{equation}
\arraycolsep=-2pt
\renewcommand{\arraystretch}{1.6}
\Gamma_1=\left[\begin{array}{cccccccc}
0&0&0&0&0&0&0&\frac{1}{z \rtt}\cr
0&\frac{q-t}{\rtt}&\ \ \frac{1}{\rtt}&0&0&0&0&0\cr
0&0&-\frac{\rtt}{q-t}&0&0&0&0&0\cr
0&0&0&0&-\frac{t}{\rtt}&0&0&0\cr
0&0&0&-\frac{\rtt}{t}&0&0&0&0\cr
0&0&0&0&0&-\frac{q-t}{\rtt}&\ \ \frac{1}{\rtt}&0\cr
0&0&0&0&0&0&\frac{\rtt}{q-t}&0\cr
\mbox{\small$z\rtt$}&0&0&0&0&0&0&0
\end{array}\right],
\end{equation}
%\comment{
\begin{equation}
\arraycolsep=1pt 
\renewcommand{\arraystretch}{1.6}
\Gamma_3=\left[\begin{array}{cccccccc}
0&0&0&\frac{1}{\sqrt{-t}\sqrt{z}}&0&0&0&0\cr
0&0&0&0&0&\frac{-q}{\sqrt{-t}\sqrt{z}}&\ \ \frac{1}{\sqrt{-t}\sqrt{z}}&0\cr
0&0&0&0&0&0&\frac{\sqrt{-t}}{q\sqrt{z}}&0\cr
\mbox{\small$\sqrt{-t}\sqrt{z}$}&0&0&0&0&0&0&0\cr
0&0&0&0&0&0&0&\frac{1}{\sqrt{-t}\sqrt{z}}\cr
0&\frac{q\sqrt{z}}{\sqrt{-t}}&\frac{\sqrt{z}}{\sqrt{-t}}&0&0&0&0&0\cr
0&0&-\frac{\sqrt{-t}\sqrt{z}}{q}&0&0&0&0&0\cr
0&0&0&0&\mbox{\small$\sqrt{-t}\sqrt{z}$}&0&0&0
\end{array}\right],
\end{equation}
\begin{equation}
\arraycolsep=-2pt
\renewcommand{\arraystretch}{1.4}
\Gamma_4=\left[\begin{array}{cccccccc}
0&0&0&0&\frac{1}{\sqrt{1-t}\sqrt{z}}&0&0&0\cr
0&0&0&0&0&\frac{1-q}{\sqrt{1-t}\sqrt{z}}&\ \ \frac{1}{\sqrt{1-t}\sqrt{z}}&0\cr
0&0&0&0&0&0&\frac{\sqrt{1-t}}{(1-q)\sqrt{z}}&0\cr
0&0&0&0&0&0&0&\frac{1}{\sqrt{1-t}\sqrt{z}}\cr
\mbox{\small$\sqrt{1-t}\sqrt{z}$}&0&0&0&0&0&0&0\cr
0&\frac{(1-q)\sqrt{z}}{\sqrt{1-t}}&\ \ -\frac{\sqrt{z}}{\sqrt{1-t}}&0&0&0&0&0\cr
0&0&\frac{\sqrt{1-t}\sqrt{z}}{1-q}&0&0&0&0&0\cr
0&0&0&\mbox{\small$\sqrt{1-t}\sqrt{z}$}&0&0&0&0
\end{array}\right]. 
\end{equation}
%}
We remark that 
the matrices $\Gamma_k$ $(k=1,3,4)$ 
are expressed as 
\begin{equation}\label{eq:Rexp} 
\begin{array}{l}
\smallskip
\Gamma_1=D(\ba_1)\exp\big(\frac{1}{q-t}E_2\big)z^{-\varpi_1}C_1,\cr
\smallskip
\Gamma_3=D(\ba_2)\exp\big(\frac{1}{q}E_2\big)z^{-\varpi_3}C_3,\cr
\Gamma_4=D(\ba_3)\exp\big(\frac{1}{q-1}E_2\big)z^{-\varpi_4}C_4,
\end{array}
\end{equation}
where we have used the notation  
\begin{equation}
D(\ba)=\diag{a_1,a_2,a_3,a_4,a_4^{-1},a_3^{-1},a_2^{-1},a_1^{-1}}
\end{equation}
for a vector $\ba=(a_1,a_2,a_3,a_4)$, $a_j\ne 0$. 
In \eqref{eq:Rexp}, $z^{\varpi_k}$ denote 
the following diagonal matrices 
associated with the fundamental weights of $\mathfrak{so}(8)$: 
\begin{equation}
\begin{array}{l}
\smallskip
z^{\varpi_1}=z^{H(1,0,0,0)}=D(z,1,1,1),\cr
\smallskip
z^{\varpi_2}=z^{H(1,1,0,0)}=D(z,z,1,1),\cr
\smallskip
z^{\varpi_3}=z^{H(\frac{1}{2},\frac{1}{2},\frac{1}{2},-\frac{1}{2})}
=D(z^{\frac{1}{2}},z^{\frac{1}{2}},z^{\frac{1}{2}},z^{-\frac{1}{2}}),
\cr
z^{\varpi_4}=z^{H(\frac{1}{2},\frac{1}{2},\frac{1}{2},\frac{1}{2})}
=D(z^{\frac{1}{2}},z^{\frac{1}{2}},z^{\frac{1}{2}},z^{\frac{1}{2}}). 
\end{array}
\end{equation}
The matrices $C_k$ are essentially permutation matrices; 
with the notation of permutation matrices 
$S_\sigma=\pmatrix{\delta_{\sigma(i),j}}_{i,j=1}^8$ for 
$\sigma\in \BS_8$,
\begin{equation}
\begin{array}{l}
\smallskip
C_1=S_{(18)(45)},\cr
\smallskip
C_3=D(1,-1,1,-1)\,S_{(14)(26)(37)(58)},\cr
C_4=D(1,-1,1,-1)\,S_{(15)(26)(37)(48)}. 
\end{array}
\end{equation}
\comment{
\begin{equation}
C_1=\left[\matrix{
0&0&0&0&0&0&0&1\cr
0&1&0&0&0&0&0&0\cr
0&0&1&0&0&0&0&0\cr
0&0&0&0&1&0&0&0\cr
0&0&0&1&0&0&0&0\cr
0&0&0&0&0&1&0&0\cr
0&0&0&0&0&0&1&0\cr
1&0&0&0&0&0&0&0
}\right],
\end{equation}
\begin{equation}
C_3=\left[\matrix{
0&0&0&1&0&0&0&0\cr
0&0&0&0&0&-1&0&0\cr
0&0&0&0&0&0&1&0\cr
-1&0&0&0&0&0&0&0\cr
0&0&0&0&0&0&0&-1\cr
0&1&0&0&0&0&0&0\cr
0&0&-1&0&0&0&0&0\cr
0&0&0&0&1&0&0&0
}\right],
\end{equation}
\begin{equation}
C_4=\left[\matrix{
0&0&0&0&1&0&0&0\cr
0&0&0&0&0&-1&0&0\cr
0&0&0&0&0&0&1&0\cr
0&0&0&0&0&0&0&-1\cr
-1&0&0&0&0&0&0&0\cr
0&1&0&0&0&0&0&0\cr
0&0&-1&0&0&0&0&0\cr
0&0&0&1&0&0&0&0
}\right],
\end{equation}
}
The matrices $D(\ba_1), D(\ba_3), D(\ba_4)$ in \eqref{eq:Rexp}
are defined by
\begin{equation}
\begin{array}{l}\smallskip
\ba_1=\left(
\frac{1}{\sqrt{t(t-1)}},\frac{q-t}{\sqrt{t(t-1)}},
-\frac{\sqrt{t(t-1)}}{q-t},\frac{-t}{\sqrt{t(t-1)}}
\right),
\cr
\ba_3=\left(
\frac{1}{\sqrt{-t}},\frac{q}{\sqrt{-t}},
\frac{\sqrt{-t}}{q},\mbox{\small$-\sqrt{-t}$}
\right),
\cr
\ba_4=\left(
\frac{1}{\sqrt{1-t}},\frac{q-1}{\sqrt{1-t}},
\frac{\sqrt{1-t}}{1-q},\frac{-1}{\sqrt{1-t}}
\right).
\end{array}
\end{equation}
We also remark that, for each $k=1,3,4$, 
the adjoint action of the 
matrix $z^{-\varpi_k}C_k$ induces 
the automorphism of the loop algebra 
$\mathfrak{so}(8)[z,z^{-1}]$ 
corresponding to the diagram automorphism 
$\sigma_1=(01)(34)$, $\sigma_3=(03)(14)$ or 
$\sigma_4=(04)(13)$, respectively.  
The remaining part of $\Gamma_k$ concerns the normalization 
of the matrices $M$ and $B$.
\par\medskip
Note that the system of differential equations \eqref{eq:Lax}
has a regular singularity at $z=0$ with exponents 
$\pm\varepsilon_j$ ($j=1,2,3,4$), and an irregular singularity at 
$z=\infty$.   
Assuming that $\varepsilon_j$ are generic, 
let us take a fundamental system of solutions 
$\Psi=\Psi(z,t)$ of \eqref{eq:Lax} with normalization 
such that 
\begin{equation}
\Psi(z,t)=\sum_{n=0}^\infty \Psi_n(t)\,z^{-H(\varepsilon)+n}
\end{equation}
around $z=0$, and that $\Psi_0(t)$ is upper triangular; 
such a $\Psi$ is determined up to the multiplication 
of constant diagonal matrices.  
Then, for each $k=0,1,2,3,4$, the B\"acklund transformation 
$s_k$ is interpreted as the transformation 
\begin{equation} 
\Psi\mapsto \widetilde{\Psi}=G_k \Psi S_k
\end{equation}
of the fundamental system of solutions, 
where $S_k=\exp(-E_k)\exp(F_k)\exp(-E_k)$ denote a lift of 
$s_k$ to the loop group of $SO(8)$. 
Similarly, for each $k=1,3,4$, 
the B\"acklund transformation $r_k$ is interpreted as the transformation
\begin{equation}
\Psi\mapsto \widetilde{\Psi}=\Gamma_k \Psi R_k, 
\end{equation}
where $R_k=z^{-\varpi_k}C_k$. 

\begin{remark}\rm\  
The system of differential equations \eqref{eq:Lax} can be 
equivalently rewritten into a chain of systems of 
rank 2.  We first extend the indexing set for $\psi_i$ and 
$\varepsilon_i$ to $\BZ$  by imposing the periodicity condition
\begin{equation}
\psi_{i+8}=z \,\psi_i ,\quad
\varepsilon_{i+8}=\varepsilon_i-1,
\quad 
\varepsilon_{9-i}=-\varepsilon_{i}
\qquad(i\in\BZ). 
\end{equation}
Then \eqref{eq:Lax} is equivalent to a system 
for the 2-vectors 
$\vec{\psi}_i=(\psi_{2i},\psi_{2i+1})^{\mbox{\scriptsize t}}$ 
($i\in\BZ$) in the following form: 
\begin{equation}
(z\partial_z+M_i)\vec{\psi}_i+N_i \vec{\psi}_{i+1}=0,
\quad 
(-\partial_t+A_i)\vec{\psi}_i+B_i\vec{\psi}_{i+1}=0,
\end{equation}
where $M_i$, $N_i$, $A_i$, $B_i$ are $2\times2$ matrices 
whose entries depend only on $t$.  
This system is formally transformed into 
\begin{equation}\label{eq:phi}
(\lambda+M_i)\vec{\varphi}_i+N_i \vec{\varphi}_{i+1}=0,
\quad 
(-\partial_t+A_i)\vec{\varphi}_i+B_i\vec{\varphi}_{i+1}=0,
\end{equation}
where $\vec{\varphi}_i=\vec{\varphi}_i(\lambda,t)$, 
by the change of coordinates $z=e^w$ and the Laplace 
transformation $\partial_w\leftrightarrow \lambda$, 
$w\leftrightarrow -\partial_\lambda$.  
Noting that $N_i$ are invertible, one can rewrite \eqref{eq:phi}
as
\begin{equation}\label{eq:chain}
\partial_t\vec{\varphi}_{i}=U_i\vec{\varphi}_i,\quad
\vec{\varphi}_{i+1}=W_i\vec{\varphi}_i
\quad(i\in\BZ),
\end{equation}
where
\begin{equation}
U_i=A_i-B_iN_i^{-1}(\lambda+M_i),\quad
W_i=-N_i^{-1}(\lambda+M_i)
\quad(i\in\BZ). 
\end{equation}
This type of $2\times 2$ nonlinear chains is investigated 
in \cite{A} in relation to Painlev\'e equations. 
It is not clear yet, however, how our system \eqref{eq:chain} 
can be related to the one employed there for obtaining 
$\Psix$.
\end{remark}

\par\bigskip
In this paper, we have presented a new Lax pair for the 
sixth Painlev\'e equation in the framework of 
the loop algebra $\mathfrak{so}(8)[z,z^{-1}]$ 
of type $D^{(1)}_4$.  
We also explained how the affine Weyl group symmetry 
of $\Psix$ can be obtained from the linear problem. 
We expect that the Lax pair discussed in this paper 
could be applied as well to 
other problems concerning Painlev\'e equations. 
Also, it would be an important problem 
to understand properly the relationship of our representation 
with various approaches to the sixth Painlev\'e equation
as in \cite{C}.

\end{document}